\documentclass{article}

\usepackage{amssymb}
\usepackage{amsmath}
\begin{document}
\begin{center}
\title*{\large\bf Integrable inhomogeneous Lakshmanan-Myrzakulov equation}\\
\author*{\small
 K.R. Esmakhanova$^a$,  G.N.
Nugmanova$^{b,}$\footnote{Corresponding author, E-mail:
cnlpgnn@satsun.sci.kz}, Wei-Zhong Zhao$^{c,}$\footnote{E-mail:
zhaowz100@163.com}, Ke
Wu$^{c,}$\footnote{ E-mail: wuke@mail.cnu.edu.cn}\\
$^a$Institute of Mathematics, Almaty 050010,
Kazakhstan\\
$^b$Institute of Physics and Technology, Almaty 050032,
Kazakhstan\\
$^c$ Department of Mathematics, Capital Normal University, Beijing
100037, China \\
} \vskip 0.2cm
\end{center}
\begin{center}
\begin{minipage}{140mm}
\hspace{0.5cm} {\small {\bf Abstract.} The integrable
inhomogeneous extension of the Lakshmanan-Myrzakulov equation is
constructed by using the prolongation structure theory. The
corresponding  L-equivalent counterpart is also given,  which is
the (2+1)-dimensional generalized NLSE.
}\\
{\small PACS:  02.30.Ik, 02.40.Hw, 75.10.Hk}\\
{\small KEYWORDS: inhomogeneous Heisenberg ferromagnet model,
(2+1)-dimensional integrable equation, motion of space curve,
Myrzakulov-I equation, Lakshmanan-Myrzakulov equation, }
\end{minipage}
\end{center}

\section{Introduction}

As an important subclass of  integrable nonlinear differential
equations, the  Heisenberg ferromagnet equation (HFE),
$\mathbf{S}_t=\mathbf{S}\times\mathbf{S}_{xx}$,  and its
(2+1)-dimensional integrable extensions have been paid more
attention [1-10]. Recently, a new (2+1)-dimensional integrable
extension was introduced by Lakshmanan and Myrzakulov (see, e.g.
Ref. [11]),
\begin{eqnarray}
{\bf S}_t&=&\{{\bf S}\times(\alpha{\bf S}_x+\beta{\bf S}_y)+u{\bf
S}\}_x,\nonumber\\
u_x&=&-\beta{\bf S}\cdot({\bf S}_x\times{\bf S}_y),\label{eq:LME}
\end{eqnarray}
where ${\bf S}=(S_{1}, S_{2}, S_{3})$ is the spin vector,
$\varepsilon S_{1}^{2}+\varepsilon S_{2}^{2}+S_{3}^{2}=1$ (for
compact case  $\varepsilon=1$ and  for noncompact case
$\varepsilon=-1$), $u$ is the real scalar function, $\alpha,
\beta$ are real constants.  When $\alpha=0, \beta=1$,
Eq.(\ref{eq:LME}) reduces to the Myrzakulov-I (M-I) equation (see,
e.g. Refs. [5-10]),
\begin{eqnarray}
{\bf S}_t&=&\{{\bf S}\times {\bf S}_y+u{\bf S}\}_x,\nonumber\\
u_x&=&-{\bf S}\cdot({\bf S}_x\times{\bf S}_y).
\end{eqnarray}
In Ref.[12], it was shown that Lakshmanan-Myrzakulov equation
(LME) (1)  is L-equivalent (about our conditional notations, see
e.g. Refs. [9-10]) and gauge equivalent to the following
integrable (2+1)-dimensional nonlinear Schr\"{o}dinger equation
(NLSE)
\begin{eqnarray}
iq_t-\alpha q_{xx}-\beta
q_{xy}-vq=0, \nonumber\\
ip_t+\alpha p_{xx}+\beta p_{xy}+vp=0, \nonumber\\
v_x=2[\alpha(pq)_x+\beta(pq)_y],
\end{eqnarray}
where $p=\varepsilon \bar{q}$.

As known, some integrable systems admit integrable inhomogeneous
extensions [13-14]. More recently, the integrable inhomogeneous
Myrzakulov-I (M-I) equation has been constructed in Ref.[14]. Note
that M-I equation is just a special case of (\ref{eq:LME}).  Thus
an interesting question that naturally arises is whether there is
more general integrable inhomogeneous extension of (\ref{eq:LME}).
The aim of this paper is to construct such integrable
inhomogeneous extension of (1) and explore some of its properties.

\section{The  integrable inhomogeneous  LME}
Let us consider the following inhomogeneous  LME,
\begin{eqnarray}
{\bf S}_t&=&\{{\bf S}\times(f_{1}{\bf S}_x+\beta{\bf S}_y)+u{\bf
S}\}_x+f_{2}{\bf S}_{x},\nonumber\\
u_x&=&-\beta{\bf S}\cdot({\bf S}_x\times{\bf S}_y),\label{eq:ILME}
\end{eqnarray}
where $\beta$ should be the constant as in Ref.[14], $f_1$ and
$f_2$ which are the functions of $(x,t)$ need to be determined. In
order to construct the integrable inhomogeneous  LME, we shall
analyze Eq.(\ref{eq:ILME}) by using the prolongation structure
theory as done in Ref.[15]. Let us first consider the prolongation
structure of (\ref{eq:ILME}) for the case of ${\bf S}_t=0$.
Setting $\mathbf{W}=\mathbf{S_x}$, $\mathbf{T}=\mathbf{S_y}$ and
taking $\mathbf{S}$,$\mathbf{T}$,$\mathbf{W}$, and u as the new
independent variables, we can define the following set of two
forms,
\begin{eqnarray}
\alpha_a&=&dS_a\wedge dx-T_ady\wedge dx,\nonumber\\
\alpha_{a+3}&=&dS_a\wedge dy-W_adx\wedge dy,\nonumber\\
\alpha_{a+6}&=&\beta(\mathbf{W}\times\mathbf{T})_adx\wedge
dy+\beta(\mathbf
{S}\times d\mathbf{T})_a\wedge dy+S_adu\wedge dy \nonumber\\
&+&uW_adx\wedge dy +f_1(\mathbf {S}\times d\mathbf{W})_a\wedge dy
+f_{1x}(\mathbf
{S}\times \mathbf{W})_a dx\wedge dy+f_2(\mathbf{W})_adx\wedge dy,\nonumber\\
\alpha_{10}&=&du\wedge
dy+\beta\mathbf{S}\cdot(\mathbf{W}\times\mathbf{T})dx\wedge dy,\nonumber\\
\alpha_{a+10}&=&dT_a\wedge dy+dW_a\wedge dx,\nonumber\\
\alpha_{14}&=&S_adW_a\wedge dy+W_adS_a\wedge dy,\nonumber\\
\alpha_{15}&=&(\mathbf{T}\cdot\mathbf{W})dx\wedge dy+S_a\cdot
dT_a\wedge dy,
\end{eqnarray}
where $a=1,2,3$, such that they constitute a closed ideal
$I=\{\alpha_i, i=1,2,\cdots,15\}$. Then we extend the above ideal
I by adding to it a set of one forms,
\begin{equation}
\Omega^k=d\xi^k+F^k(x, y, \mathbf{S},\mathbf{T},\mathbf{W},u)\xi^k
dx+G^k(x, y, \mathbf{S},\mathbf{T},\mathbf{W},u)\xi^k dy,\ \ \ \
k=1,2,\cdots ,n, \label{eq:omega}
\end{equation}
where $\xi^k$ is prolongation variable. In terms of the
prolongation condition, $d\Omega^k\subset\{I, \Omega^k\}$, we
obtain the following set of partial differential equations for
$F^k$ and $G^k$,
\begin{eqnarray}
& &\frac{\partial F^k}{\partial T_a}=\frac{\partial F^k}{\partial
u}=0, \ \ \ \ \frac{\partial G^k}{\partial
W_a}-\frac{f_1}{\beta}\frac{\partial G^k}{\partial T_a}=0
,\nonumber\\
&-&\frac{\partial F^k}{\partial S_a}T_a+\frac{\partial
G^k}{\partial S_a}W_a-\frac{\partial G^k}{\partial
u}\beta\mathbf{S}\cdot(\mathbf{W}\times\mathbf{T})
+(\frac{\partial G^k}{\partial T_a}-\frac{\partial F^k}{\partial
W_a})
\Big\{\big[\mathbf{S}\times(\mathbf{W}\times\mathbf{T})\big]_a\nonumber\\
&-&S_a(\mathbf{T}\cdot\mathbf{W})+\frac{u}{\beta}(\mathbf{S}\times\mathbf{W})_a
-\frac{f_1}{\beta}(\mathbf{W}\cdot\mathbf{W})S_a-\frac{f_{1x}}{\beta}W_a{}\Big\}
-[F,G]^k+\frac{\partial G^k}{\partial x}-\frac{\partial
F^k}{\partial y}=0,\label{eq:pde1}
\end{eqnarray}
where $[F,G]^k\equiv\sum_{l=1}^{n}F^l\frac{\partial G^k}{\partial
y^l}-\sum_{l=1}^{n}G^l\frac{\partial F^k}{\partial y^l}$. By
solving (\ref{eq:pde1}), we have the following solution,
\begin{equation}
F=\lambda\sum_{i=1}^{3}S_iX_i,\ \ \ \
G=(-\lambda\frac{f_1}{\beta}+\frac{f_2}{\beta}+\frac{u}{\beta})
\sum_{i=1}^{3}S_iX_i
+\frac{f_1}{\beta}\sum_{i=1}^{3}(\mathbf{S}\times\mathbf{W})_iX_i
+\sum_{i=1}^{3}(\mathbf{S}\times\mathbf{T})_iX_i,\label{eq:SFG}
\end{equation}
where  $X_i$, $i=1,2,3$, depend only on the prolongation variables
$\xi^k$ and have the commutation relation of the $su(2)$ Lie
algebra.

Now let us define a set of 3-form $\overline{\alpha}_i$ as
follows,
\begin{eqnarray}
\overline{\alpha}_a&=&dS_a\wedge dx\wedge dt-T_ady\wedge dx\wedge
dt,\nonumber\\
\overline{\alpha}_{a+3}&=&dS_a\wedge dy\wedge dt-W_adx\wedge
dy\wedge dt, \nonumber\\
\overline{\alpha}_{a+6}&=&\beta(\mathbf{W}\times\mathbf{T})_adx\wedge
dy\wedge dt+\beta(\mathbf {S}\times d\mathbf{T})_a\wedge dy\wedge
dt+S_adu\wedge dy\wedge dt+uW_adx\wedge dy\wedge dt\nonumber\\
&+&f_1(\mathbf {S}\times d\mathbf{W})_a\wedge dy\wedge dt
+f_{1x}(\mathbf {S}\times \mathbf{W})_a dx\wedge dy\wedge
dt+f_2(\mathbf{W})_adx\wedge
dy\wedge dt-dS_a\wedge dx\wedge dy, \nonumber\\
\overline{\alpha}_{10}&=&du\wedge
dy\wedge dt+\beta\mathbf{S}\cdot(\mathbf{W}\times\mathbf{T})dx\wedge dy\wedge dt,\nonumber\\
\overline{\alpha}_{a+10}&=&dT_a\wedge dy\wedge dt+dW_a\wedge dx\wedge dt,\nonumber\\
\overline{\alpha}_{14}&=&S_adW_a\wedge dy\wedge dt+W_adS_a\wedge dy\wedge dt,\nonumber\\
\overline{\alpha}_{15}&=&(\mathbf{T}\cdot\mathbf{W})dx\wedge
dy\wedge dt+S_a\cdot dT_a\wedge dy\wedge dt,
\end{eqnarray}
where $a=1,2,3$, such that they constitute a closed ideal. When
these two forms are null, we recover (\ref{eq:ILME}). Then we
introduce the following two forms,
\begin{equation}
\overline\Omega^k=\Omega^k\wedge dt+H^k_j\xi^j dx\wedge
dy+(A^k_jdx+B^k_jdy)\wedge d\xi^j, \ \ \ \ k=1,2,\cdots ,n,
\label{eq:somega}
\end{equation}
where the matrices of A and B  depend on the variables (x, y, t)
and the form of $\Omega^k$ is given by (\ref{eq:omega}), in which
$\lambda$ depends on the variables (x, y, t).  It is easily shown
that
\begin{eqnarray}
d\overline\Omega^k&=&\sum_{i=1}^{15}g^{ki}\overline{\alpha}_i+\sum_{j=1}^{n}\zeta_j^k\wedge\overline\Omega^j,
\end{eqnarray}
provided that the matrix H is given by
\begin{eqnarray}
H=GA-FB+A_y-B_x+A_tB-B_tA \label{eq:H}
\end{eqnarray}
and
\begin{eqnarray}
dH\wedge dx\wedge dy -\frac{1}{\beta}\frac{\partial G}{\partial
T_a}(\mathbf{S}\times\mathbf{dS})_a\wedge dx\wedge dy -\lambda_y
S_aX_adx\wedge dy\wedge dt\nonumber\\
(-\frac{(\lambda f_1)_x}{\beta}+\frac{f_{2x}}{\beta})
S_aX_adx\wedge dy\wedge dt -A_tGdx\wedge dy\wedge dt+B_tFdx\wedge
dy\wedge dt=0 .\label{eq:SH}
\end{eqnarray}
Substituting the expressions (\ref{eq:SFG}) of F and G  into
(\ref{eq:H}) and (\ref{eq:SH}), we obtain
\begin{eqnarray}
A=0, \ \ \ \  B=\frac{1}{\beta\lambda}I,
\end{eqnarray}
and
\begin{eqnarray}
\lambda_t=-\beta\lambda\lambda_y-\lambda^2f_{1x}+\lambda f_{2x}, \
\ \ \ \lambda_x=0.\label{eq:sp}
\end{eqnarray}
From Eq.(\ref{eq:sp}), we find that $f_i$, $i=1,2$, should take
the following expressions
\begin{eqnarray}
f_{i}=\mu_{i}(t)x+\nu_{i}(t).\label{eq:coeff}
\end{eqnarray}
By restricting (\ref{eq:somega}) on the solution manifold, we
obtain the Lax representation of equation (\ref{eq:ILME})
\begin{eqnarray}
\xi_x&=&-F|_{X_i=-\frac{i}{2}\sigma_i}
\xi=\frac{i\lambda}{2}\sum_{i=1}^3 S_i\sigma_i\xi,\nonumber\\
\xi_t&=&-\frac{1}{B}\xi_y -\frac{1}{B}
G|_{X_i=-\frac{i}{2}\sigma_i}\xi \nonumber\\
&=&-\beta\lambda\xi_y+\frac{i\lambda}{2}\sum_{i=1}^3[ (-\lambda
f_1+f_2+u)S_i\sigma_i +f_1(\mathbf{S}\times\mathbf{S}_x)_i\sigma_i
+\beta(\mathbf{S}\times\mathbf{S}_y)_i\sigma_i]\xi.
\end{eqnarray}
where  $\sigma_i, i=1,2,3$, are Pauli matrices, $f_1$ and $f_2$
are given by (\ref{eq:coeff}), and the spectral parameter
satisfies the nonlinear equation (\ref{eq:sp}).

It is interesting to note that the inhomogeneous LME
(\ref{eq:ILME}) admits the following integrable reductions:

i) When $\beta=0$,  it reduces to the inhomogeneous HFE [13]
\begin{eqnarray}
{\bf S}_t={\bf S}\times(f_{1}{\bf S})_{xx}+f_{2}{\bf S}_{x}.
\end{eqnarray}
It should be pointed out that we may also get this reduction by
taking $\partial_{x}=\partial_{y}$ and making some simple
transformations in (\ref{eq:ILME}).

ii) When $f_{1}=0$ and $\beta=1$, it reduces to the inhomogeneous
M-I equation [14],
\begin{eqnarray}
{\bf S}_t&=& \{{\bf S}\times{\bf S}_y+u{\bf S}\}_x+f_{2}{\bf
S}_{x},\nonumber\\
u_x&=&-{\bf S}\cdot({\bf S}_x\times{\bf S}_y).\label{eq:IM-I}
\end{eqnarray}

\section{L-equivalent  and gauge equivalent  counterpart}
In order to find the L-equivalent counterpart (about our
conditional  notations, see e.g. Refs. [9-10]) of the
inhomogeneous LME (\ref{eq:ILME}), we now consider the space curve
in 3-dimensional space $E^{3}$ with the arclength $x$, the
curvature $\kappa$ and the torsion $\tau$. Let ${\bf e}_{1}, {\bf
e}_{2} $ and ${\bf e}_{3}$ are the unit tangent, normal and
binormal vectors of a curve, respectively. Then we have  the
following  equations [6,8],
\begin{eqnarray}
{\bf e}_{jx}={\bf A}\times {\bf e}_{j}, \quad {\bf e}_{jy}={\bf
B}\times {\bf e}_{j}, \quad {\bf e}_{jt}={\bf C}\times {\bf
e}_{j},
\end{eqnarray}
where
\begin{eqnarray}
{\bf A}=\tau{\bf e}_{1}+k {\bf e}_{3}=(\tau,0,k),  {\bf
B}=\gamma_{1}{\bf e}_{1}+\gamma_{2}{\bf e}_{2}+\gamma_{3}{\bf
e}_{3}=(\gamma_{1},\gamma_{2},\gamma_{3}),  {\bf C}=\omega_{1}{\bf
e}_{1}+\omega_{2}{\bf e}_{2}+\omega_{3}{\bf
e}_{3}=(\omega_{1},\omega_{2},\omega_{3}),
\end{eqnarray}
and
\begin{eqnarray}
{\bf A}_{t}-{\bf C}_{x}+{\bf A}\times{\bf C}=0, \quad {\bf
A}_{y}-{\bf B}_{x}+{\bf A}\times{\bf B}=0,\quad {\bf B}_{t}-{\bf
C}_{y}+{\bf B}\times{\bf C}=0.
\end{eqnarray}
By taking $ {\bf S}\equiv {\bf e}_1 $ and using (21), we have
\begin{eqnarray}
{\bf B}=(\gamma_{1}, \gamma_{2},
\gamma_{3})=(\frac{1}{\beta}u+\partial_x^{-1}\tau_y,
\frac{1}{\beta \kappa}u_x,
\partial_x^{-1}(\kappa_y-\frac{\tau}{\beta\kappa}u_x)),\nonumber \\
{\bf C}=(\omega_{1}, \omega_{2},
\omega_{3})=(\frac{f_{1}}{\kappa}\kappa_{xx}+\frac{\beta}{\kappa}\kappa_{xy}-f_{1}\tau^{2}+f_{2}\tau
-\beta\tau\partial^{-1}_{x}\tau_{y},
-f_{1}\kappa_{x}-\beta\kappa_{y},
f_{2}\kappa-f_{1}\kappa\tau-\beta\kappa\partial^{-1}_{x}\tau_{y}).
\end{eqnarray}
Substituting  (23) into (22), we obtain the following equations
for curvature and torsion
\begin{eqnarray}
\kappa_t=(f_{2}\kappa-f_{1}\kappa\tau-\beta\kappa\partial^{-1}_{x}\tau_{y})_{x}-f_{1}\kappa_{x}\tau
-\beta\kappa_y\tau,
\nonumber\\
\tau_t=[\frac{f_{1}}{\kappa}\kappa_{xx}+\frac{\beta}{\kappa}\kappa_{xy}-f_{1}\tau^{2}+f_{2}\tau
-\beta\tau\partial^{-1}_{x}\tau_{y}]_{x}-f_{1}\kappa\kappa_{x}-\beta\kappa\kappa_{y}.
\end{eqnarray} In terms of the complex function
\begin{eqnarray}
q=\frac{\kappa}{2}e^{-i\partial_x^{-1}\tau},
\end{eqnarray}
we may rewritten (24) as
\begin{eqnarray}
iq_t-(f_1q)_{xx}-\beta q_{xy}-i(f_2q)_x-vq=0,\nonumber\\
ip_t+(f_1p)_{xx}+\beta p_{xy}-i(f_2p)_x+vp=0,\nonumber\\
v_x=2[(f_1pq)_x+\beta(pq)_y],
\end{eqnarray}
that is  (2+1)-dimensional NLSE. Here   $p=\varepsilon\bar q$ and
$\varepsilon=1$ ($\varepsilon=-1$) corresponds to the focusing
(defocusing ) case. The Lax representation of (26) is given by
\begin{eqnarray}
\Phi_x&=&U\Phi, \nonumber\\
\Phi_t&=&\lambda\beta\Phi_y+V\Phi,
\end{eqnarray}
where
\begin{eqnarray}
U=\frac{i\lambda}{2}+G, \quad G=\left(\begin{array}{cc}
0 & q \\
p & 0
\end{array}\right),\quad
V=\frac{i\lambda^2}{2}f_1\sigma_3+\frac{i\lambda}{2}f_2\sigma_3+\lambda
V_1 +V_0,
\end{eqnarray}
in which
\begin{eqnarray}
V_1=f_1G, \quad V_0=\left(\begin{array}{cc}
if_1pq+i\beta\partial_x^{-1}(pq)_y& f_2q-i\beta q_y-i(f_1q)_x \\
f_2p+i\beta p_y+i(f_1p)_x& -[if_1pq+i\beta\partial_x^{-1}(pq)_y]
\end{array}\right),
\end{eqnarray}
and the spectral parameter $\lambda(y,t)$ satisfies  the equation
(15). Eq. (26) also admits  two reductions:

i) When $\beta=0$,  it reduces to the inhomogeneous
(1+1)-dimensional NLSE
\begin{eqnarray}
iq_t-(f_1q)_{xx}-i(f_2q)_x-vq=0,\nonumber\\
ip_t+(f_1p)_{xx}-i(f_2p)_x+vp=0,\nonumber\\
v_x=2(f_1pq)_x,
\end{eqnarray}

ii) When $f_{1}=0$ and $\beta=1$,  it reduces to the inhomogeneous
(2+1)-dimensional NLSE
\begin{eqnarray}
iq_t- q_{xy}-i(f_2q)_x-vq=0,\nonumber\\
ip_t+ p_{xy}-i(f_2p)_x+vp=0,\nonumber\\
v_x=2(pq)_y.
\end{eqnarray}
Note that the gauge equivalence between (4) and (26) can also be
established by taking the following transformation between the
solutions of systems (17) and (27)
\begin{eqnarray} \Psi=g^{-1}\Phi,
\end{eqnarray}
where $g=\Phi|_{\lambda=0}$.

\section{Conclusion and remarks}
In this paper, we have constructed an integrable inhomogeneous
extension of (\ref{eq:LME}) and given the corresponding
L-equivalent counterpart which is the (2+1)-dimensional
generalized NLSE. It should be noted that the homogeneous HFE
admits several integrable extensions in 2+1 dimensions, such as

1$^{0}$. The Myrzakulov-VIII (M-VIII) equation  [5]
\begin{eqnarray}
{\bf S}_t&=&{\bf S}\times{\bf S}_{xx}+u{\bf S}_x,\nonumber\\
u_y&=&{\bf S}\cdot({\bf S}_x\times{\bf S}_y),
\end{eqnarray}

2$^{0}$.  The Ishimori  equation [4]
\begin{eqnarray}
{\bf S}_t={\bf S}\times({\bf S}_{xx}+\alpha^{2}{\bf
S}_{yy})+u_{x}{\bf S}_y+u_{y}{\bf S}_x,\nonumber\\
u_{xx}-\alpha^{2}u_{yy}=-2\alpha^{2}{\bf S}\cdot({\bf
S}_x\times{\bf S}_y),
\end{eqnarray}

3$^{0}$. The Myrzakulov-IX (M-IX) equation [5]
\begin{eqnarray}
{\bf S}_t&=&{\bf S}\times M_{1}{\bf S}+A_{1}{\bf S}_y+A_{2}{\bf
S}_x,\nonumber\\
M_{2}u&=&2\alpha^{2}{\bf S}\cdot({\bf S}_x\times{\bf S}_y).
\end{eqnarray}
Whether all of these equations admit inhomogeneous extensions  is
still a question for the future.
\\
\vspace{1cm} \\
{\bf Acknowledgements}

We are very grateful to Prof. R. Myrzakulov for his interest and
helpful discussions. This work is partially supported by Beijing
Jiao-Wei Key project (KZ200310028010), NSF projects (10375038
and 90403018).\\

\end{document}